\documentclass[a4paper,UKenglish,cleveref,autoref,thm-restate]{lipics-v2021}

\usepackage{tikz}
\usetikzlibrary{positioning}
\usetikzlibrary{arrows.meta}
\usetikzlibrary{shapes}
\usetikzlibrary{shapes.geometric}

\usepackage{tabularx}
\usepackage{multirow}
\usepackage{longtable}

\title{How Many Interviews Are Enough in a Software Engineering Study? Preliminary Findings on Sample Size and Saturation}

\titlerunning{How Many Interviews Are Enough in a Software Engineering Study?}

\Copyright{Ronnie de Souza Santos and Italo Santos and Mauricio Rodrigues Lima and Cleyton Magalhães}

\author{Ronnie de Souza Santos}
{University of Calgary, Canada}
{ronnie.desouzasantos@ucalgary.ca}
{0000-0003-3235-6530}
{}

\author{Italo Santos}
{University of Hawaii at Manoa, USA}
{isantos3@hawaii.edu}
{0000-0002-7545-6104}
{}

\author{Mauricio Rodrigues Lima}
{University of Hawaii at Manoa, USA}
{lima25@hawaii.edu}
{}
{}

\author{Cleyton Magalhães}
{Universidade Federal Rural de Pernambuco (UFRPE), Brazil}
{cleyton.vanut@ufrpe.br}
{0009-0005-3051-7232}
{}

\authorrunning{de Souza Santos et al.}



\EventEditors{Robert Feldt, Maria Paasivaara, Daniel Mendez, Stefan Wagner, and Marvin Mu\~{n}oz Bar\'{o}n}
\EventNoEds{5}
\EventLongTitle{20th International Symposium on Empirical Software Engineering and Measurement (ESEM 2026)}
\EventShortTitle{ESEM 2026}
\EventAcronym{ESEM}
\EventYear{2026}
\EventDate{October 8--9, 2026}
\EventLocation{Munich, Germany}
\EventLogo{}
\SeriesVolume{394}
\ArticleNo{60}

\category{Emerging Results, Vision \& Reflection Track Paper}

\begin{document}

\maketitle

\begin{abstract}
\textbf{Background.} Interview-based studies are widely used in empirical software engineering to investigate human, organizational, and socio-technical phenomena, yet interview sample adequacy and saturation are reported inconsistently across the literature. \textbf{Aims.} This paper investigates how interview sample adequacy and saturation are reported in empirical software engineering research. \textbf{Method.} We analyzed 427 papers published between 2016 and 2025 across major software engineering venues, focusing on interview sample sizes, saturation discussions, and sample adequacy justifications. \textbf{Results.} Preliminary findings indicate substantial variation in interview sample sizes, ranging from highly specialized small-sample studies to broader investigations involving large interview datasets. Studies involving 12 or fewer interviewees were common and frequently associated with specialized industrial contexts or constrained organizational access. However, the most common range was 13 to 24 interviewees, suggesting that moderate-sized samples represent the most common configuration in empirical software engineering research. Saturation and sample adequacy justifications were heterogeneous, with many studies relying on implicit or contextual reasoning rather than explicit methodological discussion. \textbf{Conclusions.} Our findings provide empirical insights into methodological reporting practices in interview-based software engineering research and contribute to ongoing discussions on qualitative rigor and transparency.
\end{abstract}

\vspace{-5px}
\section{Introduction}
\label{sec:introduction}
\vspace{-5px}

Qualitative research is an established component of empirical software engineering, particularly for investigating human, organizational, behavioral, and socio-technical aspects of software development~\cite{seaman1999qualitative, dybaa2011qualitative, lenberg2024qualitative}. Interview-based studies (e.g., case studies, surveys, and grounded theory) are widely used to investigate developer experiences, team interactions, organizational practices, decision-making processes, and perceptions that quantitative measures alone cannot capture~\cite{seaman1999qualitative, hove2005experiences}. Interviews therefore play an important role in empirical software engineering because they support contextual understanding, triangulation across data sources, and exploratory investigation of socio-technical phenomena~\cite{seaman1999qualitative, hove2005experiences, lenberg2024qualitative}.

Methodological guidelines for qualitative software engineering research emphasize transparency, reflexivity, triangulation, sampling adequacy, and saturation~\cite{seaman1999qualitative, dybaa2011qualitative, stol2016grounded, lenberg2024qualitative, ralph2020empirical}. These guidelines additionally caution against applying quantitative expectations, such as statistical representativeness, to qualitative work~\cite{ralph2020empirical}. Nevertheless, previous reviews indicate that saturation is discussed inconsistently and qualitative quality criteria are reported unevenly in software engineering publications~\cite{lenberg2024qualitative}. These issues carry practical consequences for researchers. Qualitative and interdisciplinary software engineering studies continue to face challenges during peer review and publication, with interview-based work sometimes dismissed as out of scope or \textit{``too soft''}~\cite{hyrynsalmi2025not}. Reviewer concerns frequently target the number of interviewees, methodological legitimacy, and unfamiliarity with qualitative or interdisciplinary methods~\cite{hyrynsalmi2025not}. Such concerns may become particularly pronounced in studies involving marginalized populations, senior professionals, leadership roles, or specialized industrial contexts, where recruitment is inherently difficult, and smaller samples may still provide substantively meaningful insights.

The broader qualitative methodology literature has addressed these tensions more directly. While software engineering guidelines have generally avoided prescribing exact interview sample sizes~\cite{seaman1999qualitative, dybaa2011qualitative, ralph2020empirical, lenberg2024qualitative}, interdisciplinary qualitative research has discussed interview numbers, stopping criteria, and empirical saturation ranges in greater detail~\cite{guest2006many, francis2010adequate, hennink2017code, hagaman2017many}. Prior work suggests that saturation may occur in relatively few interviews within homogeneous contexts while also cautioning against rigid numerical interpretations of adequacy~\cite{guest2006many, hennink2017code, hagaman2017many, cobern2020interviewing, elmholdt2026many}. However, despite these broader methodological discussions, there is still limited empirical evidence regarding how software engineering studies operationalize and report interview sample adequacy and saturation in practice.

The goal of this paper is to investigate how interview sample size adequacy and saturation are operationalized and reported in empirical software engineering research, with the purpose of characterizing current practices, identifying reporting patterns, and providing empirical insights into methodological decisions surrounding interview-based studies. To achieve this goal, we investigate the following research question: \textit{How are interview sample size and saturation operationalized in software engineering interview studies?} To answer this question, we conducted a large-scale review of 427 software engineering papers employing interviews, characterized interview sample size practices, explored how studies justified sample adequacy, and conducted a preliminary mapping of how saturation is discussed across the field.

Our preliminary results show that interview-based software engineering studies vary substantially in sample size, ranging from single-participant studies to investigations with more than 100 interviewees. Across 427 papers, we identified 7,407 reported interviews, with an average of 17.35 participants per study. The most common range was 13 to 24 interviewees, followed closely by studies with 5 to 12 interviewees. 
Discussions of saturation and sample adequacy were heterogeneous, with many papers relying on implicit or contextual justifications rather than explicit methodological explanations. Our study contributes to an empirical characterization of interview sample-size practices in software engineering and provides initial evidence to support more transparent reporting of saturation and sample adequacy in interview-based studies.

\vspace{-5px}
\section{Background}
\label{sec:background}
\vspace{-5px}

As discussed previously, software engineering methodological guidelines have generally avoided prescribing exact interview sample sizes, instead emphasizing saturation, methodological adequacy, triangulation, reflexivity, and alignment between research goals and qualitative procedures~\cite{seaman1999qualitative, dybaa2011qualitative, lenberg2024qualitative, ralph2020empirical}. While this flexibility is methodologically principled, it can create practical difficulties for researchers justifying interview adequacy during peer review, in particular, in studies involving difficult-to-recruit populations, interdisciplinary topics, or specialized industrial contexts~\cite{hyrynsalmi2025not}.

On the other hand, qualitative methodology literature outside software engineering has engaged with these challenges more directly, offering explicit discussion of interview numbers, saturation thresholds, stopping criteria, and sample adequacy across fields such as psychology, health research, nursing, education, and the social sciences~\cite{guest2006many, francis2010adequate, dworkin2012sample, hennink2017code, hagaman2017many, braun2021saturate, bekele2022sample}. The aim of such discussions is generally practical: to improve transparency in qualitative reporting, support saturation procedures, and establish shared expectations for publication~\cite{francis2010adequate, dworkin2012sample, bekele2022sample}.

The numerical recommendations that emerge from this literature vary considerably. For relatively homogeneous samples, recommendations range from 6 to 8 interviews~\cite{bekele2022sample}, with some studies reporting saturation within the first 12 interviews and basic metatheme elements emerging within approximately 6~\cite{guest2006many}. Another commonly discussed stopping criterion consists of an initial sample of 10 interviews followed by 3 consecutive interviews without new themes~\cite{francis2010adequate}. Beyond these general thresholds, some studies distinguish between different forms of saturation: code saturation may occur within approximately 9 interviews, whereas meaning saturation may require between 16 and 24~\cite{hennink2017code}. Others report that common themes in relatively homogeneous groups may emerge within approximately 16 interviews, while cross-cultural metathemes may require between 20 and 40~\cite{hagaman2017many}.

Broader numerical ranges reflect differences in methodological design, participant heterogeneity, and analytical goals. Phenomenological studies are commonly associated with approximately 5 to 25 interviews, grounded theory studies with 20 to 30, and ethnographic studies with 30 to 60, depending on scope and heterogeneity~\cite{bekele2022sample}; publication-oriented discussions have additionally proposed 25 to 30 interviews as a practical minimum for grounded theory journal submissions~\cite{dworkin2012sample}. At the lower end, 15 interviews have been identified as the smallest acceptable qualitative sample size in some contexts, while at the upper end, studies exceeding approximately 150 interviews may become analytically difficult to manage~\cite{bekele2022sample}. 

Despite this range of guidance, several researchers caution against rigid numerical interpretations, arguing that interview adequacy ultimately depends on study scope, participant heterogeneity, data richness, analytical depth, and interpretive goals rather than fixed thresholds~\cite{cobern2020interviewing, braun2021saturate, bekele2022sample, elmholdt2026many}, yet numerical guidance and saturation-oriented recommendations continue to appear frequently in qualitative methodology literature, reflecting an ongoing tension between methodological flexibility and the practical demands of publication.

\vspace{-5pt}
\section{Method}
\label{sec:method}

\begin{figure*}[t]
\centering
\hspace{-1cm}
\includegraphics[width=0.85\textwidth]{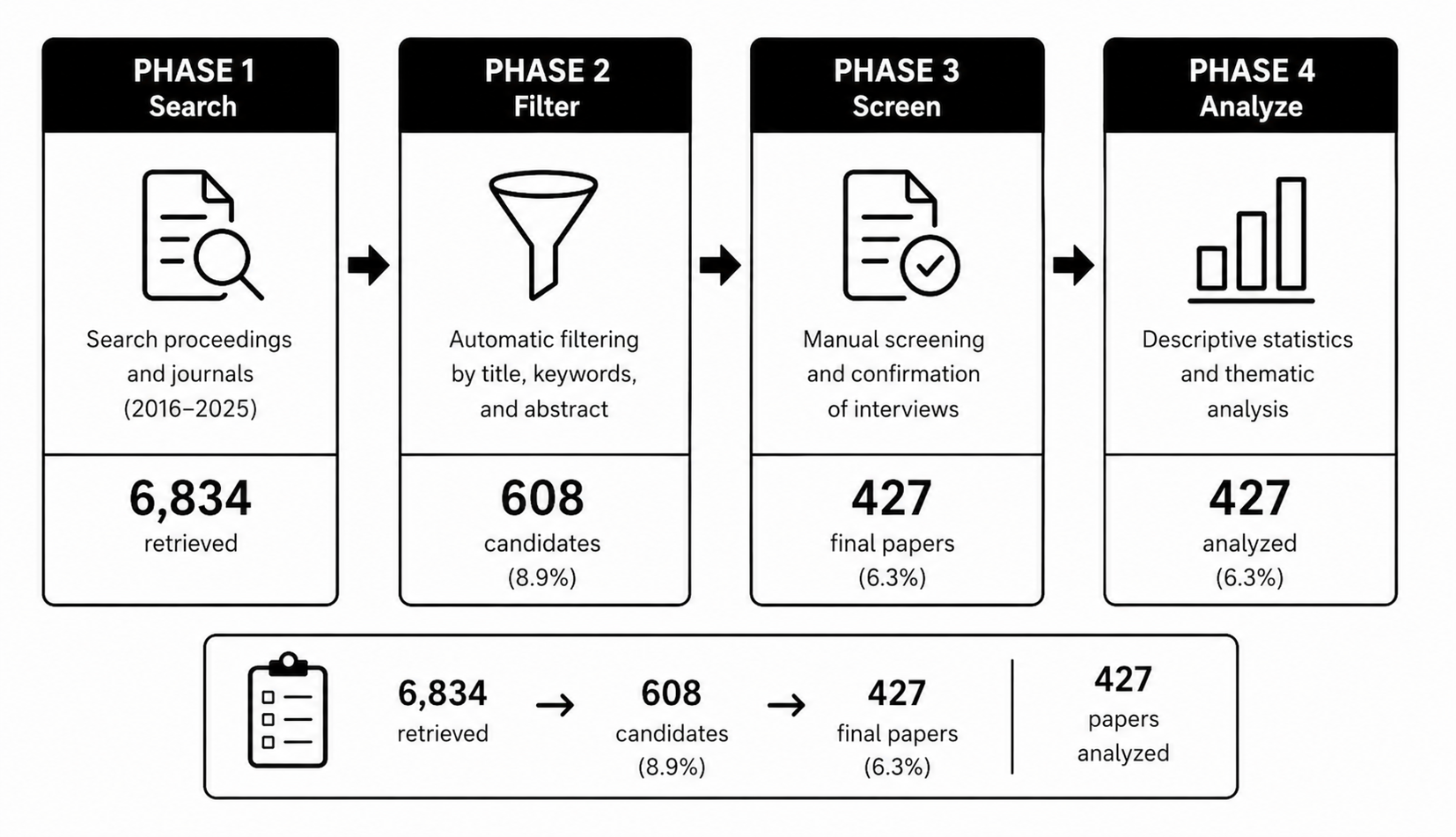}
\caption{Overview of the methodological process.}
\label{fig:method}
\vspace{-10pt}
\end{figure*}

In this study, we followed a process conceptually similar to a systematic mapping study~\cite{kitchenham2004evidence, petersen2008systematic, kitchenham2007guidelines}. We aimed to identify software engineering papers that used interviews as part of their empirical data collection process. We did not adopt an automatic search strategy because the scope of the topic was intentionally broad, namely, studies in software engineering involving interviews. Preliminary attempts indicated that conventional search strings would retrieve an excessively large and noisy corpus requiring extensive manual filtering. Therefore, we adopted a manual search strategy, which is commonly used in secondary studies in software engineering when automatic retrieval is impractical or insufficiently precise~\cite{kitchenham2007guidelines}. Our process consisted of four main stages: (1) manual identification of papers from selected software engineering venues, (2) automatic filtering of potentially qualitative studies through title, keyword, and abstract inspection, (3) manual full-text screening to confirm the actual use of interviews, and (4) preliminary descriptive and thematic analysis of the resulting corpus. Figure~\ref{fig:method} summarizes the overall methodological process adopted in this study. \\

\noindent \textbf{Search Strategy.} We manually searched papers published between 2016 and 2025 in the proceedings of the International Symposium on Empirical Software Engineering and Measurement (ESEM), the International Conference on Evaluation and Assessment in Software Engineering (EASE), the International Conference on Software Engineering (ICSE), and the ACM Joint European Software Engineering Conference and Symposium on the Foundations of Software Engineering (FSE), as well as in the Empirical Software Engineering journal (EMSE). We selected these venues because of their relevance to empirical and industrial software engineering research. ESEM, EASE, and EMSE are recognized for publishing studies with strong methodological and empirical emphasis, while ICSE and FSE were selected for their broad influence within the software engineering community and strong participation by industry practitioners, which frequently results in empirical studies involving interviews. To support the collection process, we developed a Python script to retrieve the proceedings and issues for each selected venue over the ten-year period. For conferences, we included main research tracks, industry tracks, and thematic tracks such as Software Engineering in Society (SEIS) and the Education track (SEET). We excluded co-located events and workshops. For the journal, we included both regular and special issues. This initial corpus comprised 6,834 papers. \\

\noindent \textbf{Selection Process and Data Extraction.} We then developed a second script to automatically identify papers potentially involving qualitative research methods. This process analyzed titles, keywords, and abstracts. Papers describing methods commonly associated with interview-based qualitative research, such as case studies, grounded theory, and related qualitative approaches, were automatically selected and exported to a spreadsheet for further inspection. This stage reduced the dataset from 6,834 papers to 608 candidate papers. Next, two authors manually inspected the full text of all candidate papers to confirm whether interviews had actually been conducted as part of the empirical study. We considered this manual verification step necessary because many papers referenced qualitative methods without necessarily collecting interview data. The final corpus comprised 427 papers. \\

\noindent \textbf{Data Analysis and Synthesis.} In this paper, we present emerging results identified in our analysis. So far, we have relied primarily on descriptive statistics \cite{cooksey2020descriptive} to characterize the distribution of interview sample sizes and other basic study characteristics across the identified corpus. To support interpretation, we organized interview sample sizes into characteristic ranges derived from discussions available in the qualitative research literature:

\begin{itemize}
    \item \textbf{Fewer than 5 interviews:} highly specialized, exploratory, elite, or difficult-to-access contexts~\cite{hyrynsalmi2025not}.
    
    \item \textbf{5 to 12 interviews:} homogeneous samples, focused studies, and early thematic saturation discussions~\cite{guest2006many, bekele2022sample}.
    
    \item \textbf{13 to 24 interviews:} meaning saturation and deeper interpretive analysis~\cite{hennink2017code}.
    
    \item \textbf{25 to 40 interviews:} heterogeneous, multisite, cross-cultural, and grounded theory-oriented investigations~\cite{hagaman2017many, dworkin2012sample, bekele2022sample}.
    
    \item \textbf{41 to 60 interviews:} broader ethnographic and exploratory qualitative studies involving larger datasets~\cite{bekele2022sample}.
    
    \item \textbf{More than 60 interviews:} very large qualitative datasets and discussions regarding analytical manageability and extensive qualitative coverage~\cite{bekele2022sample}.
\end{itemize}

To investigate how saturation and sample-size justification were discussed in practice, we also conducted a preliminary qualitative analysis~\cite{cruzes2011recommended}. Since this paper reports emerging results, we have not yet qualitatively analyzed the full corpus of 427 papers. Instead, we adopted an exploratory stratified selection strategy to capture variation across interview sample sizes while maintaining a manageable scope for this preliminary stage of the study. 
For this preliminary exploratory analysis, we analyzed 14 papers: 12 papers randomly selected from the interview-sample-size ranges, plus the paper with the highest number of interviews and one randomly selected paper from the two papers reporting the lowest interview count.

Random selection within each range was performed using an online randomization tool based on the paper identification codes assigned during data extraction. This strategy was adopted to obtain an initial overview of how saturation and sample size justification were discussed across studies involving substantially different interview samples, ranging from highly specialized interview settings to large qualitative datasets. The selected papers were then inspected to investigate whether and how authors discussed saturation, justified interview sample sizes, described participant accessibility, and characterized the adequacy of their interview samples. The analysis focused on methodological descriptions and reporting practices presented by the authors rather than on evaluating the objective adequacy of the interview samples themselves. Preliminary insights emerging from this exploratory analysis are discussed in this paper. In future work, we plan to extend the qualitative extraction and analysis process to the full corpus of 427 papers, enabling a broader characterization of saturation practices, reporting strategies, and approaches to sample size justification in software engineering interview studies. \\

\noindent \textbf{Threats to Validity.} \textbf{Construct Validity.} One threat to construct validity concerns the identification of papers that actually used interviews as part of their empirical data collection process. Because interview studies are described inconsistently across software engineering papers, some studies may not explicitly mention interviews in titles, keywords, abstracts, or methodological descriptions. To reduce this threat, we combined automatic filtering with manual full-text inspection conducted by two authors. Another construct-related threat concerns the categorization of interview sample size ranges. These ranges were derived from discussions available in the qualitative research literature and should not be interpreted as strict or universally accepted thresholds for adequacy or saturation. \textbf{Internal Validity.} The automatic filtering process relied on methodological terminology commonly associated with interview-based qualitative research, such as case studies, grounded theory, and surveys. Consequently, studies using interviews without explicitly describing their methods through these terms may have been missed during the automatic filtering stage. In addition, the preliminary thematic analysis reported in this paper was conducted on a reduced exploratory subset of papers rather than the full corpus. Although this strategy was intentionally adopted to support an emerging results paper, the current findings should be interpreted as preliminary and exploratory rather than exhaustive. \textbf{External Validity.} Our study focused on papers published in ESEM, EASE, ICSE, FSE, and EMSE between 2016 and 2025. Although these venues are influential within software engineering research and include substantial empirical and industry-oriented work, the findings may not generalize to all software engineering venues, related disciplines, or qualitative research communities outside software engineering. Similarly, practices associated with interview saturation and sample size justification may differ in other contexts not represented in our study. \textbf{Reliability.} Manual screening and qualitative interpretation involve researcher judgment, which may introduce inconsistencies during paper selection and thematic analysis. To reduce this threat, two authors manually reviewed the candidate papers during the interview confirmation stage and resolved conflicts using consensus meetings. We also describe our methodological process, filtering stages, and analysis procedures in detail to support transparency and reproducibility. In the complete version of this study, we plan to extend the extraction and thematic analysis process to the full corpus of 427 papers and further refine the coding process through iterative discussion among the authors.
\vspace{-5px}
\section{Results}
\label{sec:results}
\vspace{-5px}

\begin{figure*}[t]
\centering
\hspace{-1cm}
\includegraphics[width=0.85\textwidth]{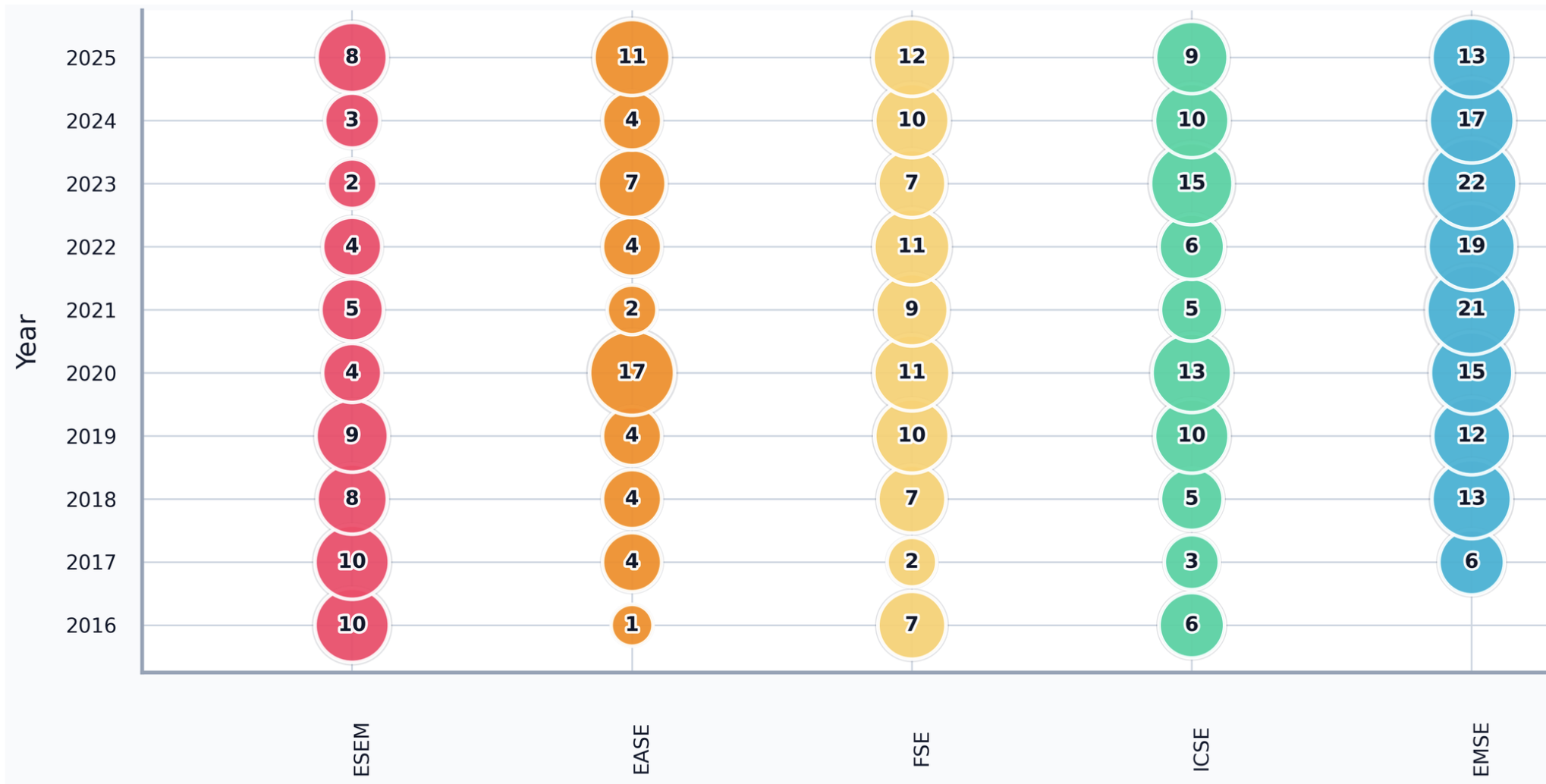}
\caption{Distribution of interview-based studies}
\label{fig:venue_year}
\vspace{-10pt}
\end{figure*}

Our sample comprised 427 software engineering papers employing interviews published between 2016 and 2025. Publication counts by year were: 24 papers in 2016, 25 in 2017, 37 in 2018, 45 in 2019, 60 in 2020, 42 in 2021, 44 in 2022, 53 in 2023, 44 in 2024, and 53 in 2025. These counts point to a rise in interview-based studies after 2018, with publication rates remaining relatively stable in subsequent years. Regarding venues, EMSE contributed the largest share (n=139), followed by FSE (n=85), ICSE ( n=83), ESEM (n=62), and EASE (n=58), indicating that interview-based empirical research is distributed across multiple major software engineering venues rather than concentrated in a single outlet. Figure~\ref{fig:venue_year} presents the combined distribution across publication years and venues.

The identified studies also varied in interview type. Semi-structured interviews were the dominant approach, appearing in 407 papers, followed by structured interviews (n=9) and unstructured interviews (n=3). We additionally identified 5 interview surveys or qualitative surveys, 2 focus group interviews, and 3 studies combining multiple interview types. Regarding interview mode, in-person interviews appeared in 82 papers, remote interviews in 58, and video-call interviews in 54. Mixed formats appeared in 65 papers and email interviews in 9. A substantial portion of the analyzed papers did not report how interviews were conducted, pointing to variability and incompleteness in methodological reporting practices.

\vspace{-5px}
\subsection{Interview Sample Size Characteristics}
\label{sec:sample_ranges}
\vspace{-5px}

\begin{figure*}[t]
\centering
\hspace{-1cm}
\includegraphics[width=0.95\textwidth]{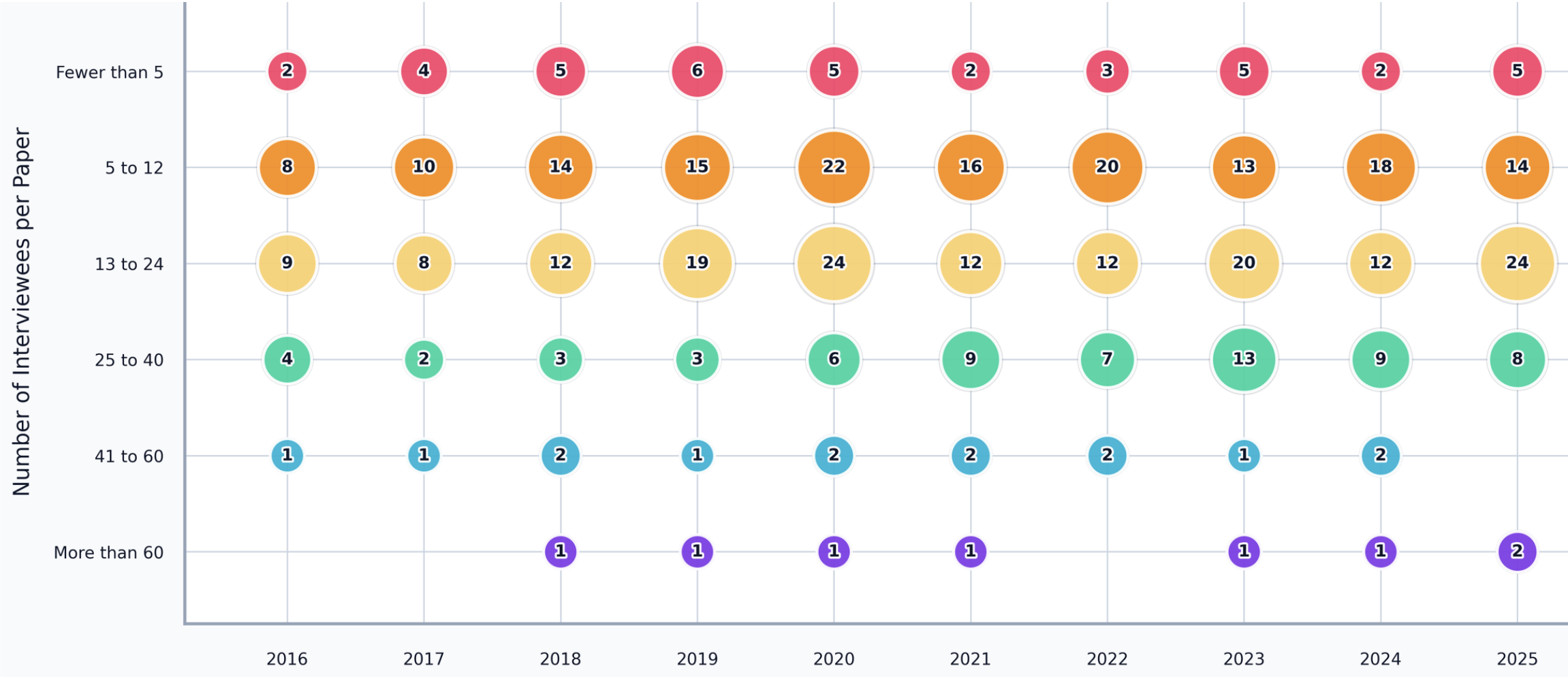}
\caption{Distribution of interview sample size ranges}
\label{fig:distribution_ranges}
\vspace{-10pt}
\end{figure*}

Interview sample sizes varied substantially across the analyzed studies. Across the 427 papers, we identified 7,407 reported interviews, with an average of 17.35 participants per study. The smallest studies relied on a single participant, while the largest reported 100 interviewees. Figure~\ref{fig:distribution_ranges} presents the distribution of interviewee counts across the analyzed papers. Studies involving 13 to 24 interviewees were the most frequent (n=152), followed closely by studies with 5 to 12 interviewees (n=150). Studies involving 25 to 40 interviewees appeared in 64 papers, while studies with fewer than 5 interviewees appeared in 39 papers. Studies involving 41 to 60 interviewees were identified in 14 papers, and studies employing more than 60 interviewees were comparatively rare (n=8).

These results indicate that interview based software engineering research spans a broad range of sample sizes rather than converging around a single dominant range. The analyzed studies cover multiple sample size ranges discussed in the qualitative methodology literature, from small homogeneous samples to moderate sized qualitative investigations and broader exploratory studies. Studies with 13 to 24 interviewees were the most prevalent, suggesting that moderate sized interview samples are common in empirical software engineering research. Studies with 5 to 12 interviewees also appeared frequently, indicating that smaller interview studies remain widely used across the field. The substantial number of studies involving 25 to 40 interviewees further suggests that larger qualitative and exploratory investigations also represent an important segment of the literature. These ranges should not be interpreted as prescriptive methodological thresholds, but rather as interpretive groupings informed by the qualitative research literature. The observed variation likely reflects differences in research objectives, participant accessibility, methodological approach, analytical scope, study context, and publication practices.

\vspace{-5px}
\subsection{Preliminary Analysis of Sample Size Justification and Saturation}
\vspace{-5px}

Among the 14 papers randomly selected for this preliminary exploratory analysis, the full list is provided in the supplemental material due to space restrictions. Authors often provided some explanation of participant selection, even when explicit methodological justification for the final sample size was limited. These explanations tended to focus on contextual, organizational, or practical recruitment considerations rather than formal qualitative sampling guidance. Explicit saturation discussions were comparatively rare, appearing in only four papers, two of which were grounded theory investigations, a methodology that relies heavily on iterative sampling and theoretical saturation. \\

\noindent \textbf{Sample Justification.} The analyzed papers showed heterogeneous approaches to describing and justifying interview sample sizes. Smaller studies typically justified participant selection by the specificity of the investigated context, the uniqueness of participants, or the limited availability of individuals directly involved in the phenomenon under study. For example, an investigation involving a single hacker participant or release managers responsible for patch uplift decisions. Studies with moderate sample sizes more commonly emphasized variation across organizational contexts, products, stakeholder groups, or technical perspectives. One study involving 21 interviews framed the work as a qualitative survey intended to capture variation across industrial settings rather than conduct an in-depth analysis of a single case. Another involving 23 interviews deliberately included API producers associated with both adoption and non-adoption of lambda support to capture contrasting perspectives within the Java ecosystem. Larger studies typically justified participant counts through broader organizational coverage, multisite designs, or heterogeneous stakeholder groups. The studies involving 42 and 75 interviews relied on grounded theory procedures with iterative recruitment and theoretical sampling. A study involving more than 100 interviews defined recruitment as part of a customer discovery process covering workflows, onboarding practices, and organizational adoption challenges across multiple professional roles. Although explicit methodological justification was often limited, most studies still provided contextual reasoning for participant selection.

\noindent \textbf{Saturation.}
Explicit discussion of saturation was infrequent in the analyzed subset. The two grounded theory studies described iterative data collection, theoretical sampling, and continuous refinement of interview questions until theoretical saturation was reached, operationalized as the point at which additional interviews no longer produced new analytical insights. Outside grounded theory, saturation was discussed less systematically. One study treated saturation as the recurrence of previously identified codes and used this observation to iteratively adapt the interview guide, rather than as a stopping criterion for recruitment. Another explicitly acknowledged that theoretical saturation was not achieved, arguing that saturation may have limited temporal validity in rapidly evolving security contexts where attacker behavior continuously changes. Explicit saturation reporting thus remains relatively uncommon in this subset of software engineering interview studies. When discussed in methodological detail, it was primarily associated with grounded theory investigations involving iterative recruitment and theoretical sampling.

\vspace{-5px}
\subsection{How are interview sample size and saturation operationalized in software engineering interview studies?}
\vspace{-5px}

Considering both the quantitative characterization of interview sample sizes and the preliminary qualitative analysis of reporting practices, our findings indicate that software engineering interview studies employ highly variable sample sizes, ranging from single-participant studies in highly specialized contexts to investigations involving more than 100 interviewees. Currently, the most common range in empirical software engineering is 13 to 24 interviewees, with an average of 17.35. However, no single interview sample size can be considered universally appropriate for software engineering interview studies. Sample sizes appear to be strongly dependent on research objectives, methodological approach, analytical depth, participant accessibility, organizational scope, and the nature of the phenomenon under investigation. Interview sample adequacy is therefore highly context-dependent rather than determined by fixed numerical expectations. Yet, researchers tend to provide only contextual explanations for participant selection, and explicit justification for why a specific number of interviews was considered sufficient remains limited. Saturation discussion, in turn, is largely confined to grounded theory investigations involving iterative recruitment and theoretical sampling. These emerging results suggest that interview sample adequacy, justification, and saturation deserve greater methodological attention in empirical software engineering research.

\vspace{-5px}
\section{Discussion}
\label{sec:discussion}
\vspace{-5px}

Our emerging results suggest that software engineering interview research partially aligns with broader qualitative research traditions while also diverging from them in important ways. Similar to qualitative research in psychology, health, education, and the social sciences, software engineering studies frequently rely on contextual reasoning when defining interview samples, with participant accessibility, organizational scope, technical specialization, and the nature of the investigated phenomenon all influencing sample adequacy decisions. This aligns with broader qualitative understandings that adequacy depends on research goals, heterogeneity, and analytical depth rather than fixed numerical thresholds~\cite{braun2021saturate, bekele2022sample, cobern2020interviewing, hagaman2017many, hennink2017code}.

At the same time, qualitative methodology literature in other fields has increasingly emphasized explicit reporting of saturation procedures, stopping criteria, and sample adequacy rationale~\cite{francis2010adequate, hennink2017code, bekele2022sample}. In contrast, our findings suggest that many software engineering studies rely primarily on implicit or contextual justification without systematically articulating why the selected number of participants was considered analytically sufficient. This divergence may partially reflect the historical development of empirical software engineering, where methodological reporting has often prioritized technical artifacts, evaluation procedures, and empirical outcomes over theorized discussion of interpretive adequacy, reflexivity, or saturation logic~\cite{hyrynsalmi2025not, baltes2022sampling}.

Organizational realities also play a role. Many empirical studies investigate industrial environments characterized by restricted access, confidentiality constraints, non-disclosure agreements, and difficult-to-recruit populations, such as senior architects, security specialists, or release managers who are directly responsible for specific technical decisions. In these contexts, recruiting even a small number of highly knowledgeable participants may require substantial effort, organizational permissions, and sustained professional relationships~\cite{hove2005experiences, baltes2022sampling}. Our findings further suggest that participant availability is frequently a binding constraint rather than a methodological choice, and sample sizes may therefore reflect what was realistically achievable rather than what was theoretically optimal. Under these conditions, contextual justification is not only pragmatic but also necessary, and evaluating such studies against numerical expectations derived from less-constrained research environments may be inappropriate. Software engineering interview research may therefore be closer to applied and organizational qualitative research, where access constraints strongly influence study design and where the depth and relevance of participant knowledge often matter more than the number of participants recruited.

Despite these realities, our findings suggest several areas for improvement. Greater transparency in sample adequacy reasoning would benefit peer review and replication without requiring fixed numerical recommendations. Software engineering research would also benefit from broader engagement with contemporary saturation concepts beyond grounded theory, including code saturation, meaning saturation, and information power~\cite{hennink2017code, braun2021saturate, malterud2016sample}, and from stronger qualitative-specific reporting expectations within existing methodological guidelines~\cite{ralph2020empirical, baltes2022sampling}. Our findings also indicate that studies involving fewer than 5 specialist participants, or between 5 and 12 interviewees, are common and methodologically legitimate in software engineering research. Small samples should not be used as grounds for paper rejection or negative reviews, provided authors offer clear justification for their sample size and, preferably, engage with saturation. The methodological quality of an interview study cannot be reduced to participant count alone~\cite{cobern2020interviewing, baltes2022sampling}.

Finally, we caution against applying sample-size recommendations from other disciplines without accounting for the contextual particularities of software engineering research. The observed variability suggests that interview adequacy depends heavily on industrial access, participant specialization, and methodological design. That said, if a reference point is needed, our findings suggest that 13 to 24 interviewees represent the most empirically common range in software engineering, and may serve as a practical starting point for researchers planning interview studies. Numerical guidance from broader qualitative literature may serve better as interpretive reference points than prescriptive requirements~\cite{cobern2020interviewing, hagaman2017many, hennink2017code}. The central challenge is not identifying a universally correct number of interviews, but improving transparency about how adequacy decisions are made and how they connect to the goals of empirical investigation.
\vspace{-5px}
\section{Conclusions and Future Work}
\label{sec:conclusion}
\vspace{-5px}

We investigated how interview sample size adequacy and saturation are operationalized in empirical software engineering research. Sample sizes vary substantially, with 13 to 24 interviewees representing the most frequent range, while studies with 12 or fewer participants were also common. Very small samples were commonly linked to specialized industrial contexts, expert populations, or constrained organizational access, and should not be dismissed on participant count alone, provided adequate justification is offered. Larger samples, by contrast, were more frequently associated with broader organizational coverage, heterogeneous participant groups, multisite investigations, and exploratory designs requiring wider contextual representation. Adequacy decisions thus appear strongly dependent on contextual, organizational, and methodological factors rather than shared numerical criteria. Saturation and explicit sample-adequacy justification were heterogeneous and comparatively uncommon outside grounded-theory studies, suggesting methodological transparency in interview-based software engineering research deserves greater attention. \textbf{Future Work.} We plan to extend the analysis to the full corpus of 427 papers, investigating how justification practices, saturation operationalization, and recruitment decisions vary across methodologies, venues, industrial contexts, and participant profiles. These steps aim to support clearer methodological guidance, improve transparency in qualitative reporting, and support more consistent, context-sensitive evaluation of qualitative studies by researchers, reviewers, and editors.

\vspace{-5px}
\section{Data Availability}
\vspace{-5px}

The data used in this study is available at: \url{https://figshare.com/s/64a69f589cdc50d804f2}

\bibliography{lipics-v2021-sample-article}

\end{document}